\documentstyle[12pt]{article}
\newcommand{\beq}{\begin{equation}}
\newcommand{\eeq}{\end{equation}}

\begin{document}
\def\lag{\langle}
\def\rag{\rangle}
\begin{titlepage}
\vspace{2.5cm}
\baselineskip 24pt
\begin{center}
\large\bf{Groundstate Properties of the 3d Ising Spin Glass$^{1}$}\\
\vspace{1.5cm}
\large{ Bernd A. Berg$^{2,3}$ , Ulrich H.E. Hansmann$^{3,4}$
and Tarik Celik$^{5}$ }
\end{center}
\vspace{3cm}
\begin{center}
{\bf Abstract}
\end{center}

We study zero--temperature properties of the 3d Edwards--Anderson Ising
spin glass on finite lattices up to size $12^3$.  Using
multicanonical sampling we generate large numbers of groundstate
configurations in thermal equilibrium. Finite size scaling with a
zero--temperature scaling exponent $y = 0.74 \pm 0.12$ describes the
data well. Alternatively, a descriptions in terms of Parisi mean field
behaviour is still possible. The two scenarios give significantly
different predictions on lattices of size $\ge 12^3$.
\vfill

\footnotetext[1]{{This research project was partially funded by the
Department of Energy under contracts DE-FG05-87ER40319, DE-FC05-85ER2500,
and by MK Research, Inc.\,.}}
\footnotetext[2]{{Department of Physics, The Florida State University,
                      Tallahassee, FL~32306, USA. }}
\footnotetext[3]{{Supercomputer Computations Research Institute,
                      Tallahassee, FL~32306, USA.}}
\footnotetext[4]{{Department of Physical Sciences,
Embry--Riddle Aeronautical University, Daytona Beach, FL 32114, USA.}}
\footnotetext[5]{{Department of Physics, Hacettepe University, Ankara,
		  Turkey.}}
\end{titlepage}

\baselineskip 24pt

The theoretical problem to determine the equilibrium groundstate
structure of spin glasses has remained an important, but elusive,
question. It is generally agreed that the statistical mechanics of
the infinitely ranged Sherrington--Kirkpatrick Ising spin glass is
essentially understood. The replica--symmetry breaking mean field (MF)
scheme discovered by Parisi \cite{Pa1} exhibits infinitely many
low--temperature states whose properties are consistent with
simulations \cite{Yo1}, see \cite{Bi1} for a recent overview.
Whether, for temperatures below the spin glass phase transition point
$T_c$, the more realistic short--ranged 3$d$ Edwards--Anderson Ising
spin glass model (EAI)  also exhibits Parisi mean field behaviour has
become a central question. It was answered in the negative by proponents
of a simple scaling ansatz [4--7]. These droplet scaling (DS) theories
suggest that no more than two pure states (related via a global flip)
exist at any temperature.
The MF approximation is surely valid for $d\to \infty$, and it has been
suggested \cite{Bray} that $d=6$ is the upper critical dimension which
separates the MF from the DS scenario.

The EAI Hamiltonian is given by
$$ H\ =\ - \sum_{<ij>} J_{ij} s_i s_j . \eqno(1) $$
Here the sum $<ij>$ goes over nearest neighbours. We consider 3$d$
systems with periodic boundary conditions and $N=L^3$ spins.
The exchange interactions $J_{ij}=\pm 1$ between $N$ spins $s_i=\pm 1$
are randomly distributed over the lattice with the constraint
$\sum_{<ij>} J_{ij} = 0$. For each system there
are $(3N)!/[(3N/2)!]^2$ realizations of the quenched random variables $J=\{
J_{ij} \}$. Recent simulations \cite{Cara} in a magnetic field favour the
mean field picture rather than the alternative droplet model.  However, it
has been pointed out that equilibrium at sufficiently low temperatures has
not been reached \cite{Fish}.

A quantity of decisive importance is the probability density $P(q)$
of the Parisi order parameter $q$:
$$ P(q) =  < P(q) >_J =
  {[(3N/2)!]^2 \over (3N)!} \sum_J P_J (q) . \eqno(2) $$
By $<\cdot >_J$ we denote averaging over the realizations $J$.
For a fixed realization $P_J (q)$ is the probability density of the
overlap
$$ q = q_J = \left. {1\over N} \sum_i s_i^1 s_i^2 \right|_J . \eqno(3) $$
Here  $s_i^1$ and $s_i^2$ denote two replica ({\it i.e} statistically
independent configurations) of the realization $J$ at temperature $T$.
Due to magnetic field zero we have the symmetry $P(-q)=P(q)$, such that
$ \int_{-1}^{+1} q^n P(q) dq = 0 $ for $n$ odd. We
therefore define averages over the range $0\le q \le 1$:
$$ \overline{q^n} = <q^n> = 2 \int_0^1 q^n P(q) dq. \eqno(4) $$
For the spin glass susceptibility $\chi_q = N <q^2>$ the MF as well as the DS
scenario suggest divergence $\chi_q \sim N$ for $T<T_c$. However they give
significantly different predictions for the variance
$$  \sigma^2(\overline{q}) = <(\overline{q}-q)^2>  . \eqno(5) $$
In the limit $N\to \infty$ one has $\sigma^2 (\overline{q}) \to finite$
in MF theory, while $\sigma^2 (\overline{q}) \sim L^{-y} \to 0$ within
the DS approach. Here $y=-y_T$ is the zero--temperature scaling exponent
\cite{Bray},
denoted $\theta$ in \cite{Huse}, which governs also the finite size
scaling (FSS) corrections of the expectation values$^1$:
$\overline{q^n}_L - \overline{q^n}_{\infty} \sim L^{-y}$ for DS,
whereas we assume 1/Volume corrections for MF theory.

\footnotetext[1]{{If necessary we indicate the lattice size by an
additional subscript $L$, which is otherwise dropped.}}

{}~~Lack of self--averaging is one prominent feature of MF behaviour.~~\ In
reference \cite{Cara} $<\int(P(q)-P_J(q))^2dq>_J$ was studied. This was
criticised by the authors of \cite{Fish}. Following their suggestion we
estimate
$$ \sigma^2_J (\overline{q^2} ) =
<(\overline{q^2}-\overline{q^2}_J)^2>_J. \eqno(6) $$
Again, this quantity stays finite in MF theory, but drops off
$\sim L^{-y}$ in the DS picture.

With the development of multicanonical techniques for disordered systems
\cite{our2} it has become feasible to generate spin glass groundstates in
thermal equilibrium; see \cite{our3} for a brief, general review, and
\cite{um} for the earlier umbrella sampling. A pilot study
for the model at hand has been presented in \cite{our4}. In
essence a multicanonical spin glass simulation proceeds in three steps.
First Monte Carlo (MC) weights are recursively constructed which will
allow to simulate an ensemble, the ``multicanonical'', which yields
canonical expectation values in the temperature range $0\le T\le\infty$
through use of the spectral density. Secondly, equilibrium
configurations with respect to the multicanonical ensemble are generated
by means of standard MC. In a third step canonical expectation values
at desired temperatures are obtained from the analysis. Multicanonical
sampling circumvents the notorious ergodicity problems of canonical
low temperature spin glass simulations through regular excursions into
the disordered phase, {\it while} staying in equilibrium.

In this paper we focus on the investigation of groundstate properties. A
lower bound on the number of statistically independent groundstates
sampled is obtained by counting how often the system moves from the
energy $E\ge 0$ region to the groundstate energy $E_{\min}$, and back to
the $E\ge 0$ region. This has been termed ``tunneling'' \cite{our2} and
we follow this notation, but
one should bear in mind that the free energy barriers are actually not
overcome by a tunneling process. With present techniques the tunneling
time approximately increases $\sim V^{3.4}$ in 3d \cite{our4}. This
slowing down limits our investigation to rather moderately sized lattices.

We have performed simulations for $L=4$, 6, 8, 12 ($N=64$, 216, 512,
1728). For $L\le 8$ the sum (2) is approximated through 512 randomly chosen
realizations of the $\{ J_{ij}\}$, whereas we have only 7 realizations
for $L=12$. For all 1,543 cases multicanonical parameters were determined
recursively. Then each system was simulated twice with independent random
starts and random numbers. This constitutes our two independent replica per
realization. In these production runs iterations were stopped when a preset
number of tunneling events $n_{\tau}$ had occurred: $n_{\tau} = 128\ (L=4)$,
64 ($L=6$), 32 ($L=8$) and 10 ($L=12$). Despite this decrease in
tunneling events, the average number of updates per spin $n_s$ (sweeps) did
steadily increase. Approximate values are: $n_s = 8\cdot 10^4\ (L=4)$,
$10^5\ (L=6)$, $7.6\cdot 10^6\ (L=8)$ and $50\cdot 10^6\ (L=12)$. The
average CPU time spent on one $L=8$ replica was approximately 800 minutes
on an IBM 320H workstation.

Per replica we have stored up to 2,048 groundstate configurations. Due to
correlations the number of encountered groundstates is, of course, much
larger than $n_{\tau}$. If the number exceeded 2,048, the stored
configurations were randomly selected from the total set. For groundstate
configuration $n$ this is elegantly done  on--line by picking a random
integer $i_r$ in the range $1\le i_r\le n$. Configuration $n$ is stored at
position $i_r$ if $i_r\le 2,048$ and discarded otherwise. For both
replica the same groundstate energy has to be found between all tunneling
counts. This is a strong, albeit not rigorous, criterium to ensure that
the correct groundstate energy has not been missed.

On a semi--log scale figure~1 depicts the thus obtained probability
densities (2) for the Parisi order parameter. The $L=12$ probability
density, presented without error bars, is very bumpy due to the small
number of realizations, and
will only be reliable for a few of the considered physical quantities.
Note, altogether the data respect the $P(q)=P(-q)$ symmetry well.
For $L=8$ figure~2 plots the $P_J(q)$ probability densities of two
rather extreme $L=8$ realizations: two peak shape versus continuous
distribution. Various different shapes in--between these extremes are
also found. Figure~3 shows all $L=8$ realizations together. From
figures~1--3 it is evident that only a careful quantitative analysis
of these distributions may give hints concerning the $L\to\infty$
groundstate distribution.

Our estimates for various measured quantities are summarized in table~1.
The error bars are with respect to the different realizations, which
are statistically independent and enter with equal weights. Table~2
summarizes two--parameter fits of the data, assuming alternatively MF
theory or the DS ansatz to be true. If MF and DS scenario lead to the
same functional form, the fit is marked ``All''. The $\infty$ column
gives the infinite volume extrapolations of the considered quantity,
$Q$ is the goodness of fit, and R$_{12}$ comments Yes or No on the
reliability of the $L=12$ data for the purposes of the particular fit.
With an exception for $P_{\max}$, the MF and ALL fits are of the form
$a_1+a_2/L^3$. In the MF as well as in the DS scenario the self--overlap
gives rise to a $\delta$--function singularity. Therefore $P_{\max}$
is supposed to grow $\sim V$ and the appropriate fit is $a_1L^3+a_2$.
Including all data points the fit is still consistent. Although, omitting
the smallest lattice indicates that $L=4$ may not fully exhibit the
asymptotic behaviour.

Let us now discuss $\sigma^2(\overline{q})$ and
$\sigma^2_J(\overline{q^2})$. The DS fit is
$a_1L^{-a_2}$. Fits and data are depicted in figure~4.
For each case we give two fit curves. The upper one relies on three
data points $(L=4,6,8)$, whereas the lower one includes also the $L=12$
result. When only three data are used, MF and DS fits are both consistent
($Q=0.10$ and $Q=0.43$). Once the $L=12$ data point is included, the
consistency of the MF fit becomes marginal $(Q=0.04)$. However, from the
$L=8$ data we have the experience that 10\% of the realizations amount
to 99\% of the $P(0)$ contribution. Consequently, the $L=12$ data
suffer not only from large statistical fluctuations, but are altogether
unreliable for quantities which are sensitive to the small $q$ distribution.
The above fits were also used for $P(0)$, but the data are too inaccurate
to yield meaningful results. We now rely on the three--point fits for
$\sigma^2(\overline{q})$ and $\sigma^2_J(\overline{q^2})$. The two $y=a_2$
estimates are still compatible, and we summarize them to $y = 0.74 \pm 0.12$.
The error bar is not reduced as both estimates rely on the same data set.

Assuming that our $L=4-8$ lattices show already
typical scaling behaviour, we conclude from figure~4 that similarly
accurate data on lattices up to size $L=16$ would discriminate
between the MF and the DS ansatz. Due to the slowing down, our new data
indicate $\sim V^{3.9}$, the
needed CPU time would be about 1,000 times larger than the one spent on the
present investigation. With upcoming massively parallel devices in the
teraflop range such a factor can be achieved.

The not yet discussed DS fits are of the form $a_1+a_2L^{-y}$ with
$y=0.74$. Let us first comment on the groundstate energy fits. The energy
is self--averaging, $L=12$ contributes reasonably
accurate results, and we rely on all data sets. Most interesting,
the DS fit is consistent, whereas the MF fit with 1/Volume
corrections is ruled out. Unfortunately, there is still a catch to it.
It may well be that the corrections to the uncritical MF behaviour are
exponentially small. Although FSS corrections for larger systems would
then be greatly reduced, the disadvantage at the present level is that
the appropriate $a_1+a_2\exp(-a_3L)$ fit has three free parameters. A
consistent ($Q=0.19$) fit is then still possible. Again,
accurate data on lattices up to $L=16$ would allow to differentiate this
behaviour from DS.

Let us remark that the groundstate entropy stays finite,
as about 5\% of the spins can still be freely flipped due to the exact
degeneracy of the $\pm 1$ quenched random variables \cite{Kirk}. Because
of this pathology of the model non--critical FSS correction seem to be
appropriate in either scenario, and 1/Volume corrections work indeed well.

A relevant consistency check for the correctness of the DS picture is that the
infinite volume estimates of $\overline{q}_{\max}$, $\sqrt{\, \overline{q^2}}$
and $\overline{q}$ have to agree. Figure~5 shows the MF and DS fits for these
quantities. For $L$ a log--scale is used to exhibit $L\to\infty$ clearly,
and the infinite volume estimates are depicted towards the end off the scale.
With the previously determined zero--temperature exponent the values are
indeed consistent. For the MF fits $\overline{q}_{\infty} <
\sqrt{\, \overline{q^2}_{\infty}} < \overline{q}^{\,\infty}_{\max}$, as it
should be then. It is notable that fitting with a wrong zero--temperature
exponent may produce entirely inconsistent results. For
instance with $y=0.2$ one finds $\overline{q} > 1 >
\overline{q}^{\,\infty}_{\max}$.

In summary, the DS ansatz is so far consistent. Our investigation
presents the first MC estimate of the zero--temperature scaling exponent.
Obviously, our lattices are too small to allow seminal results. In
particular, the MF picture is still a valid alternative. It is clear that,
either by brute computer power or by algorithmic improvements, simulations
on larger lattices will become feasible. It seems, we are approaching a
numerical conclusion about the correct groundstate picture of the $3d$
EAI model.
\hfill\break

{\bf Acknowledgements:} We would like to thank David Huse for a useful
e-mail communication. Our simulations were performed on the SCRI cluster of
fast workstations.
\hfill\break 
\hfill\break\vskip 20pt

\section*{Tables}

\begin{table}[h]
\centering
\begin{tabular}{||c|c|c|c|c|c|c|c||}                                  \hline
$      $      & $L = 4$    & $L = 6$    & $L = 8$    & $L=12$      \\ \hline
              &            &            &            &             \\ \hline
$-e^0   $     & 1.7378 (28)& 1.7674 (13)& 1.7799 (08)& 1.7936 (27) \\ \hline
$s^0    $     & 0.0740 (09)& 0.0535 (05)& 0.0479 (03)& 0.0437 (24) \\ \hline
$\sigma(e^0)$&0.00373 (25)&0.000749 (51)&0.000277 (19)&0.000057 (34)\\ \hline
$\sigma(s^0)$&0.000784 (53)&0.000190 (13)&0.0000709 (48)&0.000046 (27)\\ \hline
$\overline{q}$& 0.785 (07) & 0.800 (06) & 0.817 (06) & 0.880 (28)  \\ \hline
$\overline{q^2}$
              & 0.669 (09) & 0.685 (08) & 0.703 (07) & 0.786 (38)  \\ \hline
$\overline{q}_{\max}$
              & 0.939 (06) & 0.9252 (25)& 0.9160 (15)& 0.901 (10)  \\ \hline
$P_{\max}$    & 4.08 (15)  & 6.04 (21)  & 8.38 (26)  & 16.0 (5.0)  \\ \hline
$P (0) $      & 0.206 (30) & 0.231 (41) & 0.140 (37) & 0.007 (07)  \\ \hline
$\sigma^2(\overline{q})$
	      & 0.0532 (26)& 0.0446 (27)& 0.0354 (27)& 0.013 (13)  \\ \hline
$\sigma^2_J(\overline{q^2})$
	      & 0.0385 (17)& 0.0258 (15)& 0.0214 (15)& 0.006 (06)  \\ \hline
\end{tabular}
\caption{{\em Data.       }}
\end{table}
\hfill\break

\begin{table}[h]
\centering
\begin{tabular}{||c|c|c|c|c|c|c|c|c|c|c|c|c||}               \hline
                            &$\infty$& $ a_1 $     & $ a_2 $    &$Q$
& $ a_1 $     & $ a_2 $    &$Q$ &R$_{12}$&Fit     \\ \hline
                            &        &$L=4\to 8$:  &            &
&$L=4\to 12$: &            &    &  &      \\ \hline
$-e^0$                      & $a_1$  & 1.78541 (92)&$-3.23~(21)$ &0.01
& 1.78637 (88)&$-3.23~(21)$&$10^{-5}$&Y&MF      \\ \hline
$-e^0$                      & $a_1$  & 1.84051 (42)&$-0.279~(18)$&0.16
& 1.8389 (40) &$ 0.274 (16)$&0.29&Y&DS      \\ \hline
$s^0$                       & $a_1$  & 0.04419 (46)& 1.947 (97)  &0.49
& 0.04413 (46)& 1.954 (97) &0.64&Y&All     \\ \hline
$L^3 \sigma^2(e^0)$         & $a_1$  & 0.128 (10)  & 7.1 (1.4)   &0.94
& 0.128 (10)  & 7.2 (1.4)   &0.85&Y&All     \\ \hline
$L^3 \sigma^2(s^0)$         & $a_1$  & 0.0352 (25) & 0.98 (31)   &0.56
& 0.0353 (25) & 0.97 (31)  &0.55&Y&All     \\ \hline
$\sigma^2(\overline{q})$    & $a_1$  & 0.0359 (25) & 1.15 (25)   &0.10
& 0.0349 (25) & 1.22 (26)  &0.04&N&MF      \\ \hline
$\sigma^2(\overline{q})$    &  $0$   & 0.116 (24)  & 0.55 (13)   &0.43
& 0.123 (25)  & 0.59 (12)  &0.28&N&DS      \\ \hline
$\sigma^2_J(\overline{q^2})$& $a_1$  & 0.0195 (15) & 1.23 (16)   &0.59
& 0.0188 (14) & 1.28 (16)  &0.06&N&MF      \\ \hline
$\sigma^2_J(\overline{q^2})$&  $0$   & 0.129 (25)  & 0.88 (12)   &0.44
& 0.137 (26)  & 0.92 (12)  &0.29&N&DS      \\ \hline
$\overline{q}$              & $a_1$  & 0.8168 (53)& $-2.18~(62)$&0.20
& 0.8196 (51)& $-2.41~(62)$&0.01&N&MF      \\ \hline
$\overline{q}$              & $a_1$  & 0.863 (17) &$-0.225~(61)$&0.44
& 0.873 (16) &$-0.258~(59)$&0.09&N&DS      \\ \hline
$\overline{q^2}$            & $a_1$  & 0.7029 (64) &$-2.32~(78)$&0.20
& 0.7059 (63) &$-2.58~(78)$&0.02&N&MF      \\ \hline
$\overline{q^2}$            & $a_1$  & 0.752 (21) &$-0.239~(76)$&0.48
& 0.764 (20) &$-0.278~(74)$&0.08&N&DS      \\ \hline
$\overline{q}_{\max}$       & $a_1$  & 0.9135 (20)&  1.82 (42)  &0.12
& 0.9129 (19)&  1.90 (42)  &0.09&Y&MF      \\ \hline
$\overline{q}_{\max}$       & $a_1$  & 0.8808 (85)&  0.165 (36) &0.73
& 0.8791 (81)&  0.171 (34) &0.77&Y&DS      \\ \hline
$P_{\max}$                  &$\infty$& 3.58 (17)   &0.00974 (67)&0.04
& 3.60 (17)   &0.00958 (65)&0.08&Y&All     \\ \hline
$P_{\max}$ & $\infty$      &      &$L=6$,8,12: &$\to$
& 4.39 (39)   &0.0077 (11) &0.66&Y&All     \\ \hline
$P (0)   $                  &  $a_1$ & 0.168 (35)  &  2.8 (3.3) &0.13
& 0.002 (07)  & 14.9 (2.1) &$10^{-6}$&N&MF  \\ \hline
$P (0)   $                  &  $0$   & 0.36 (22)   & 0.38 (36)  &0.17
& 0.76 (39)   & 1.21 (24)  &$10^{-11}$&N&DS  \\ \hline
\end{tabular}
\caption{{\em Fits.   }}
\end{table}
\hfill\break

\section*{Figure Captions}
\vspace{10pt}
\hspace{5pt}

\noindent
{\bf Fig. 1} \hspace{5pt}
Probability densities $P(q)$ for the Parisi order parameter
($L=4, 6, 8$ and 12).
\vspace{10pt}

\noindent
{\bf Fig. 2} \hspace{5pt}
Probability densities $P_J(q)$ for two very different $L=8$ realizations.
\vspace{10pt}

\noindent
{\bf Fig. 3} \hspace{5pt}
All $P_J(q)$ probability densities for $L=8$.
\vspace{10pt}

\noindent
{\bf Fig. 4} \hspace{5pt}
Fits for $\sigma^2(\overline{q})$ (Data q) and
$\sigma^2_J(\overline{q^2})$ (Data q2).
\vspace{10pt}

\noindent
{\bf Fig. 5} \hspace{5pt}
Fits and extrapolations for $\overline{q}_{\max}$ (up),
$\sqrt{\,\overline{q^2}}$ (middle) and $\overline{q}$ (down).
\vspace{10pt}


\vskip 30pt

\begin{thebibliography}{99}

\bibitem{Pa1} M. M\'ezard, G. Parisi, and M.A. Virasoro, {\it Spin
	      Glass Theory and Beyond} (World Scientific, 1987).

\bibitem{Yo1} A.P. Young, Phys. Rev. Lett. {\bf 51}, 1206 (1983).

\bibitem{Bi1} A.P. Young, J.D. Reger, and K. Binder, in {\it The Monte
Carlo Method in Condensed Matter Physics}, edited by K. Binder, Topics
in Applied Physics, Vol. 71 (Springer, New York, 1992).

\bibitem{Drop} W.L. McMillan, J. Phys. C {\bf 17}, 3179 (1984).

\bibitem{Bray} A.J. Bray and M.A. Moore, in {\it Heidelberg Colloquium on
                Glassy Dynamics}, edited by J.L. van Hemmen and I. Morgenstern,
                Lecture Notes in Physics, Vol. 275 (Springer, New York, 1987).

\bibitem{Huse} D.A. Fisher and D.A. Huse, Phys. Rev. B {\bf 38}, 386 (1988).

\bibitem{Juerg} A. Bovier and J. Fr\"ohlich, J. Stat. Phys.
		{\bf 44}, 347 (1986).

\bibitem{Cara} S. Caracciolo, G. Parisi, S. Patarnello and N. Sourlas,
               J. Phys. France {\bf 51}, 1877 (1990).

\bibitem{Fish} D.A. Fisher and D.A. Huse, J. Physique I France {\bf 1},
	       612 (1991).

\bibitem{our2} B. Berg and T. Celik, Phys. Rev. Lett. {\bf 69}, 2292 (1992).

\bibitem{our3} B. Berg, Int. J. Mod. Phys. C {\bf 3}, 311 (1992).

\bibitem{um} G.M. Torrie and J.P. Valleau, J. Comp. Phys. {\bf 23}, 187 (1977).

\bibitem{our4} B. Berg, T. Celik and U. Hansmann, Europhys. Lett. {\bf 22},
	       63 (1993).

\bibitem{Kirk} S. Kirkpatrick, Phys. Rev. B {\bf 16}, 4630 (1977).

\end{thebibliography}
\end{document}